\documentclass[12pt]{article}
\usepackage[utf8]{inputenc}

\usepackage{amsfonts,amssymb,amsmath,enumerate}
\usepackage{theorem,cite}
\usepackage{indentfirst}
\usepackage{textcomp}
\usepackage{bbm}
\usepackage{tikz}
\usepackage{subfigure}

\usepackage{color,ulem}

\theorembodyfont{\rmfamily}

\newcommand{\bea}{\begin{eqnarray}}
\newcommand{\eea}{\end{eqnarray}}
\newcommand{\beano}{\begin{eqnarray*}}
\newcommand{\eeano}{\end{eqnarray*}}
\newcommand{\beq}{\begin{equation}}
\newcommand{\eeq}{\end{equation}}

\newcommand{\mb}[1]{\hspace{2.1ex}\mbox{#1}\hspace{2.1ex}}
\numberwithin{equation}{section}

          \def\fc{{\mathfrak c}}

     \def\fr{{\mathfrak r}}

\def\cS{{\cal S}}

\newcommand{\II}{{\mathbb I}}

\newcommand{\MM}{{\mathbb M}}

\newcommand{\wh}[1]{\widehat{#1}}

\newcommand{\braket}[1]{\langle #1\rangle}

\begin{document}
\setcounter{page}{0}
\pagestyle{empty}
%
%
\begin{center}

 {\LARGE  {\sffamily   A stochastic model solvable without integrability} }\\[1cm]

\vspace{10mm}
  
{\Large 
F Mathieu\footnote{fabien.mathieu@lapth.cnrs.fr} and
 E. Ragoucy\footnote{eric.ragoucy@lapth.cnrs.fr}}
 \\[.41cm] 
{\large Laboratoire d'Annecy-le-vieux de Physique Th{\'e}orique LAPTh\\
 CNRS and Universit{\'e} Savoie Mont Blanc.\\[.242cm]
 BP 110, F-74941  Annecy-le-Vieux Cedex, 
France. }
\end{center}
\vfill

\begin{abstract}
We introduce a model with diffusive and evaporation/condensation processes, depending on 3 parameters obeying some inequalities. 
The model can be solved in the sense that all correlation functions can be computed exactly without the use of integrability. 

We show that the mean field approximation is not exact in general. This can be shown by looking at the analytical expression
of the two-point correlation functions, that we provide.
We confirm our analysis by numerics based on direct diagonalisation of the Markov matrix (for small values of 
the number of sites) and also by Monte-Carlo simulations (for a higher number of sites).

Although the model is symmetric in its diffusive rates, it exhibits a left / right asymmetry driven by the evaporation/condensation processes. 
We also argue that the model can be taken as a one-dimensional model for catalysis or fracturing processes.
\end{abstract}

\vfill\vfill

\newpage
\pagestyle{plain}

\section{Introduction}

One-dimensional Markovian models of hard core interacting particles hopping on a lattice, such as the Asymmetric Simple Exclusion Process ({ASEP}), are commonly studied as solvable non-equilibrium statistical physics problems \cite{DEHP, tracy2009asymptotics}. The ASEP model is known to be integrable, which provides powerful tools to study it \cite{CRV}. Some of its variations keep the integrability property \cite{DiSSEP}, but this property is rather fragile: it does generally not survive modifications.
The matrix ansatz (or DEHP method, named from its authors) is another way to get insight on some models \cite{DEHP, sandow}, but outside the integrability framework, the elaboration of a matrix ansatz as hard as a direct diagonalization of the Markov process \cite{KS}, see also \cite{BE} for a recent review on the subject. However, asking for the stationarity of the densities (one-point correlation functions) is in general enough to get their analytical expressions. The same is true for the current, but starting with two-point correlation functions, the calculation becomes in general laborious \cite{Derrida2007}. 
Few examples are known for such calculations, but they are often done for fermions on graphs, see e.g. \cite{Eisler, Klich,GKR}, or use the mean field approximation \cite{SMW} 

In the present paper, we study a model that allows a decoupling of the $n$-point correlation functions from the higher ones.
We  take advantage of this hierarchy to compute the correlation functions step by step. We provide explicit formulae for the densities, the currents and the  2-points correlation functions and observe that the correlation functions differ from the mean field approximation. We also provide the general recursion relations that allow to compute the $n$-point correlation functions.

The model  has some characteristics of the Symmetric Simple Exclusion Process ({SSEP}) but also allows evaporation and condensation of particles on the lattice. The condensation and evaporation rates are free parameters of the model, obeying some inequalities to ensure the Markovian property.
This model possesses two distinguished features: 
\begin{itemize}
\item It catches and releases single particles and pairs of particles, the rate of the later being different from the rate of the former. This allows to see it as a one-dimensional Markovian model for fracturing or catalysis of molecules.
 \item Although the model is symmetric in its diffusive rates, the evaporation/condensation processes induce a right/left asymmetry. As a consequence,  the model has some properties characteristic of a SSEP model, but also properties closer to an ASEP one.
 \end{itemize}

This model is a priori not integrable, in the sense that it is not connected to the $R$-matrices that one can find in the literature. We come back on this point in the conclusion.

After presenting the general features of the model in section \ref{sec:model}, we derive formulae for the one-site observables in section \ref{sec:dens-curr} and explicit the catalysis by comparing the evaporation currents of single particles against the evaporation current of pairs of particles. We show what combination of the parameters determines whether the lattice bounds particles together or splits them apart. In section \ref{sec:correl-fct}, we discuss $n$-point correlation functions, giving an explicit formula for 2-point correlation functions and presenting the general recursion relations for any $n$-points correlation function. We investigate in section \ref{sec:c0=1} a special case where the model is integrable and can be solved completely.  Finally, in section \ref{sec:num}, we present numerical results and plot observables. 

\section{The model \label{sec:model}}
We study a SSEP-like model on a one-dimensional lattice where in addition to the usual symmetric interaction of particles on the lattice, 
evaporation from and condensation on the lattice can occur. 
These processes can apply to particles, and also to 'molecules' constituted of two particles sitting on two adjacent sites.
The lattice is coupled to two reservoirs, that can exchange particles in the usual way.
The rates can be divided in two parts: one corresponding to the DiSSEP model \cite{CRV}, 
where only 'molecules' can condensate or evaporate, plus a second additional part where single particles can also condensate or evaporate.
This allows us to interpret the model as a fracturing process, with 'molecules' constituted of two 'particles' on adjacent sites, condensing or evaporating from the lattice at rates which vary from the ones of single 'particles'. Hence, the one-dimensional lattice can be thought of as a 
{catalyst} immersed in a gaz of molecules, fracturing them into particles, see also discussion in section \ref{sec:fract}.

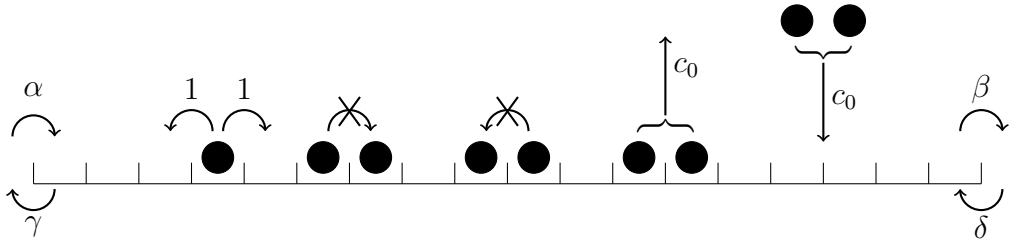
\begin{figure}[htb]
\begin{center}
 \begin{tikzpicture}[scale=0.7]
\draw (-4,0) -- (14,0) ;
\foreach \i in {-4,-3,...,14}
{\draw (\i,0) -- (\i,0.4) ;}
\draw[->,thick] (-4.4,0.9) arc (180:0:0.4) ; \node at (-4.,1.8) [] {$\alpha$};
\draw[->,thick] (-3.6,-0.1) arc (0:-180:0.4) ; \node at (-4.,-0.8) [] {$\gamma$};

\draw  (-0.5,0.5) circle (0.3) [fill,circle] {};
\draw[->,thick] (-0.6,1) arc (0:180:0.4); \node at (-1.,1.8) [] {$1$};
\draw[->,thick] (-0.4,1) arc (180:0:0.4); \node at (0.,1.8) [] {$1$};
\draw  (1.5,0.5) circle (0.3) [fill,circle] {};
\draw  (2.5,0.5) circle (0.3) [fill,circle] {};
\draw[-> ,thick] (1.6,1) arc (180:0:0.4);
\draw[thick] (1.8,1.65) -- (2.2,1.1);
\draw[thick] (2.2,1.65) -- (1.8,1.1) ;
\draw  (5.5,0.5) circle (0.3) [fill,circle] {};
\draw  (4.5,0.5) circle (0.3) [fill,circle] {};
\draw[->,thick] (5.4,1) arc (0:180:0.4); 
\draw[thick] (4.8,1.65) -- (5.2,1.1);
\draw[thick] (5.2,1.65) -- (4.8,1.1) ;

\draw  (8.5,0.5) circle (0.3) [fill,circle] {};
\draw  (7.5,0.5) circle (0.3) [fill,circle] {};
\node at (8,1.1) [rotate=-90] {$\Big{\{}$};
\draw[->,thick] (8,1.3) -- (8,2.8); \node at (8.4,2.2) [] {$c_0$};
\draw  (10.5,3.1) circle (0.3) [fill,circle] {};
\draw  (11.5,3.1) circle (0.3) [fill,circle] {};
\node at (11,2.5) [rotate=90] {$\Big{\{}$};
\draw[->,thick] (11,2.3) -- (11,0.8); \node at (11.4,1.6) [] {$c_0$};

\draw[->,thick] (13.6,1) arc (180:0:0.4) ; \node at (14.,1.8) [] {$\beta$};
\draw[->,thick] (14.4,-0.1) arc (0:-180:0.4) ; \node at (14.,-0.82) [] {$\delta$};
 \end{tikzpicture}
 \end{center}
 \caption{Dynamical rules of the model (DiSSEP part).}
 \label{fig:DiSSEP}
\end{figure}

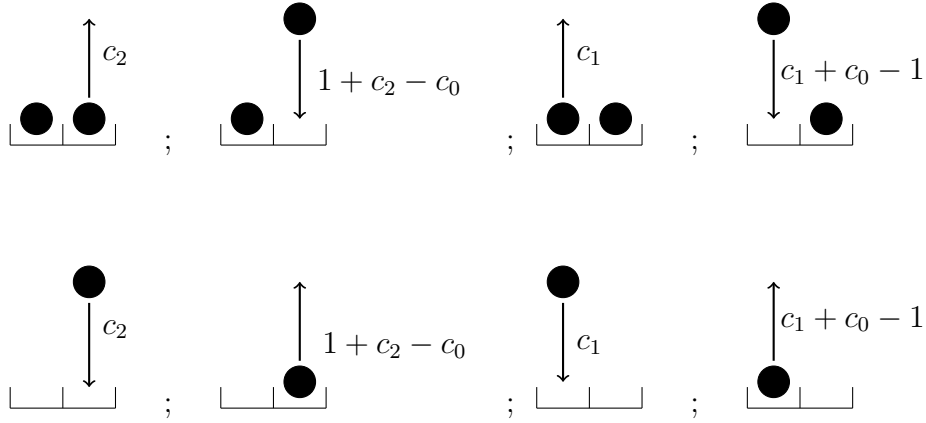
\begin{figure}[htb]
\begin{center}
\begin{tikzpicture}[scale=0.7]
\draw (0,0) -- (2,0) ;
\foreach \i in  {0,1,2} 
{\draw (\i,0) -- (\i,0.4) ;}
\draw  (1.5,2.4) circle (0.3) [fill,circle] {};
\draw[->,thick] (1.5,2.) -- (1.5,0.4); \node at (2.,1.5) [] {$c_2$};

\node at (3,0)[] {;}; 
\draw (4,0) -- (6,0) ;
\foreach \i in {4,5,6}
{\draw (\i,0) -- (\i,0.4) ;}
\draw  (5.5,0.5) circle (0.3) [fill,circle] {};
\draw[->,thick] (5.5,0.9) -- (5.5,2.4); \node at (7.3,1.2)  []  {$1+c_2-c_0$};

\node at (9.5,0)[] {;};
 \draw (10,0) -- (12,0) ;
\foreach \i in {10,11,12}
{\draw (\i,0) -- (\i,0.4) ;}
\draw  (10.5,2.4) circle (0.3) [fill,circle] {};
\draw[->,thick] (10.5,2) -- (10.5,0.5); \node at (11.,1.2)  []  {$c_1$};

\node at (13.,0)[] {;};
\draw (14,0) -- (16,0) ;
\foreach \i in {14,15,16}
{\draw (\i,0) -- (\i,0.4) ;}
\draw  (14.5,0.5) circle (0.3) [fill,circle] {};
\draw[->,thick] (14.5,0.9) -- (14.5,2.4); \node at (16,1.7) [] {$c_1+c_0-1$};


\draw (0,5) -- (2,5) ;
\foreach \i in {0,1,2}
{\draw (\i,5) -- (\i,5.4) ;}
\draw  (1.5,5.5) circle (0.3) [fill,circle] {};
\draw  (0.5,5.5) circle (0.3) [fill,circle] {};
\draw[->,thick] (1.5,5.9) -- (1.5,7.4); \node at (2.,6.7)  []  {$c_2$};

\node at (3.,5)[] {;};
\draw (4,5) -- (6,5) ;
\foreach \i in {4,5,6}
{\draw (\i,5) -- (\i,5.4) ;}
\draw  (5.5,7.4) circle (0.3) [fill,circle] {};
\draw  (4.5,5.5) circle (0.3) [fill,circle] {};
\draw[->,thick] (5.5,7) -- (5.5,5.5); \node at (7.2,6.2) [] {$1+c_2-c_0$};

\node at (9.5,5)[] {;};
 \draw (10,5) -- (12,5) ;
\foreach \i in {10,11,12}
{\draw (\i,5) -- (\i,5.4) ;}
\draw  (10.5,5.5) circle (0.3) [fill,circle] {};
\draw  (11.5,5.5) circle (0.3) [fill,circle] {};
\draw[->,thick] (10.5,5.9) -- (10.5,7.4); \node at (11.,6.7)  []  {$c_1$};

\node at (13.,5)[] {;};
\draw (14,5) -- (16,5) ;
\foreach \i in {14,15,16}
{\draw (\i,5) -- (\i,5.4) ;}
\draw  (14.5,7.4) circle (0.3) [fill,circle] {};
\draw  (15.5,5.5) circle (0.3) [fill,circle] {};
\draw[->,thick] (14.5,7) -- (14.5,5.5); \node at (16,6.4) [] {$c_1+c_0-1$};

 \end{tikzpicture}
 \end{center}
 \caption{Dynamical rules of the model (additional transitions for single particles).}
 \label{fig:additionnal}
\end{figure}

The values of the different rates are shown in figures \ref{fig:DiSSEP} and \ref{fig:additionnal}. 

It leads to the following local Markov matrices.
In the bulk:
\beq\label{eq:Markovmat}
M=\left(\begin{array}{cccc}
-c_0- c_1-c_2 & 1+c_2-c_0 & c_1+c_0-1 & c_0 \\ 
c_2 & -1- c_1-c_2 & 1 & c_1 \\ 
c_1 & 1 & -1- c_1-c_2 & c_2 \\ 
c_0 & c_1+c_0-1 & 1+c_2-c_0 &-c_0- c_1-c_2
\end{array}\right)\,.
\eeq
The off-diagonal terms of $M$ are positive when 
\beq\label{eq:cont}
0\leq c_k\,,\ k=0,1,2\mb{and}1-c_1\leq c_0\leq 1+c_2.
\eeq
Note that these constraints are automatically satisfied if one chooses $c_1=c_0=c_2\geq\frac12$. It leads to a one-parameter model
that could be worth studying.

On the left boundary, we have
\beq
B=\left(\begin{array}{cc}
-\alpha & \gamma \\  \alpha & -\gamma 
\end{array}\right).
\eeq

On the right boundary, we have
\beq
\bar B=\left(\begin{array}{cc}
-\delta & \beta \\  \delta & -\beta 
\end{array}\right).
\eeq

The full Markovian matrix takes the form
\beq\label{eq:marko-full}
\MM=B_1+\sum_{j=1}^{L-1} M_{j,j+1} +\bar B_L.
\eeq

\subsection{General properties}
\subsubsection*{Particle/hole symmetry}
The exchange of particles and holes corresponds to a conjugation
\beq 
\begin{cases}
M_{j,j+1}\to V\otimes V\,M_{j,j+1}\,V\otimes V= M_{j,j+1}\\[1ex]
 B\to V\,B\,V \mb{;} \bar B\to V\,\bar B\,V
\end{cases}
\mb{with} V=\left(\begin{array}{cc} 0 & 1 \\  1 & 0 \end{array}\right)\,.
\eeq
Remark that, in the bulk, the model is invariant under this transformation: only the boundaries break it.
It corresponds to the exchange $\alpha\ \leftrightarrow\ \gamma$ and $\beta\ \leftrightarrow\ \delta.$

\subsubsection*{Transposition in the bulk}
The transposition of the matrix $M$ is still a Markovian matrix. 
To connect the matrix $M^T$ with a matrix of the form \eqref{eq:Markovmat}, one needs to reverse the orientation of the lattice, which amounts to conjugate the matrix by the permutation matrix 
\[
	P = \begin{pmatrix}
		1 & 0 & 0 & 0 \\
		0 & 0 & 1 & 0 \\
		0 & 1 & 0 & 0 \\
		0 & 0 & 0 & 1
	\end{pmatrix}\,.
\]
Then, the matrix $M'=P\,M^T\,P$ is of the form \eqref{eq:Markovmat} with parameters
\beq 
c'_0=c_0\,,\quad c'_1=1+c_2-c_0 \mb{and} c'_2=c_1+c_0-1,
\eeq
obeying $c'_j\geq0$ and $1-c'_1\leq c'_0\leq 1+c'_2$.

When $c_0=1$, $M$ is self-transposed and $P$ is not needed, since it commutes with $M$. 
In that case, the model is simpler and the mean field approximation becomes exact, see  section \ref{sec:c0=1}.

\subsubsection*{Comparison with DiSSEP}
A SSEP model with dissipation, called DiSSEP, has already been studied in \cite{CRV,DiSSEP}. 
Its local Markov matrix reads:
\beq
W=\left(\begin{array}{cccc}
-\kappa & 0 & 0 & \kappa \\ 
0 & -1 & 1 & 0\\ 
0 & 1 & -1 & 0 \\ 
\kappa & 0 &  0 &-\kappa
\end{array}\right).
\eeq
The matrices $W$ and $M$ coincide only when the parameters take the values
  $\kappa=c_0=1$ and $c_1=c_2=0$, which corresponds to the free fermions point, see a detailed study in \cite{DiSSEP}. 
We will show that for the present model, $c_0=1$ (with $c_1$ and $c_2$ kept free) is enough to simplify the presentation,  see section \ref{sec:c0=1}.

Note also that
we checked that for generic values of their parameters it is \underline{not} possible to relate the two matrices 
$M$ and $W$ by transformations of the form
\beq\label{eq:conjug}
M=U\otimes U\,W\,U^{-1}\otimes U^{-1}+T\otimes\II_2-\II_2\otimes T+\lambda\,\II_4\,,
\eeq
where $U$ is an arbitrary invertible $2\times 2$ matrix,  $T$ an arbitrary $2\times 2$ matrix corresponding to telescoping terms, and $\lambda$ a free parameter.

\subsubsection*{Case of periodic lattice}
In the case of a periodic lattice, where the site $L+1$ is identified with the first site, the model is at equilibrium.
Indeed, the stationary state can be computed, it takes the form
\beq
|\cS_0\rangle =\begin{pmatrix} 1\\1\end{pmatrix}\otimes \ldots \otimes \begin{pmatrix} 1\\1\end{pmatrix}.
\eeq
The densities are constant, $\braket{n_j}=\frac12$, $\forall\ j$, and there is no current.
This shows the similarity between the present model and the SSEP model: it is the presence of reservoirs that brings the system out of equilibrium.

\section{Currents and densities\label{sec:dens-curr}}
As we shall see, when asking for the stationarity of the densities and currents, we get recursion relations involving 
densities only. There is a decoupling from the higher correlation functions, such that the recursion can be solved.
This property extends to higher correlation functions, see section \ref{sec:correl-fct}.

To simplify the presentation, we introduce the ratios:
\beq \label{eq:Lambda}
\Lambda=\frac{\gamma-\alpha}{2c_0+2c_1+\gamma+\alpha}\,,\quad 
\wh\Lambda = \frac{\beta-\delta}{2+2c_2+\beta+\delta} 
\mb{and}
\Gamma=\frac{1-c_0}{1+c_0+c_1+c_2}.
\eeq
The parameter $\Lambda$ (resp. $\wh\Lambda$) encodes the dependence in the right (resp. left) boundary, while $\Gamma$ corresponds
to the site (i.e. the position)
 dependence for densities and currents 
(see below). When $\Gamma=0$ or 1, there is no dependence in the position (except for the first and the last sites). 

Remark that $\Gamma$ is a decreasing function of $c_0\geq0$ with positive parameters $c_1$ and $c_2$.
Without any constraint, the extreme values of $\Gamma$ are obtained for $c_0\to \infty$ and $c_1$, $c_2$ finite 
(which gives $\Gamma=-1$) and for $c_0=0$ with $c_1=c_2=0$ (which provides $\Gamma=1$). 
However, due to the constraints \eqref{eq:cont}, when  $c_0\to \infty$, we must take 
 $c_2\to \infty$, while to get $c_0=0$ we must choose $c_1=1$. This leads to
$$-\frac12\leq\Gamma\leq\frac12.$$
One can define two sub-classes of models that lead to the extremal values for $\Gamma$:
\begin{itemize}
\item One can take $c_2=c_0-1$. The inequalities \eqref{eq:cont} are then satisfied for $c_0\geq1$. 
Then, one gets $\Gamma=\frac{1-c_0}{2c_0+c_1}$, and the limit $c_0\,\to\,\infty$ (keeping $c_1$ finite) gives $\Gamma=-\frac12$. Up to a scaling factor $c_0$, the limit provides a model with no free parameter and no diffusion. Its local Markov matrix reads
\beq
\lim_{c_0\to\infty}\Big(\frac1{c_0}M_{c_2=c_0-1} \Big)= \left(\begin{array}{cccc}
-2 & 0 & 1 & 1 \\ 
1 & -1 & 0 & 0 \\ 
0 & 0 & -1 & 1 \\ 
1 & 1 & 0 &-2
\end{array}\right)\,.
\eeq
\item One can take $c_1=1-c_0$ and $c_2={\fc_2}{c_0}$. The inequalities \eqref{eq:cont} are then satisfied for $c_0\leq1$. 
Then, one gets $\Gamma=\frac{1-c_0}{2+c_0\fc_2}$, and the value $c_0=0$ (keeping $\fc_2$ finite)
gives $\Gamma=\frac12$. Its local Markov matrix at $c_0=0$ has no free parameter and reads:
\beq
M'= \left(\begin{array}{cccc}
-1 & 1 & 0 & 0 \\ 
0 & -2 & 1 & 1 \\ 
1 & 1 & -2 & 0 \\ 
0 & 0 & 1 &-1
\end{array}\right)\,.
\eeq
\end{itemize}

\subsection{Densities}
The time evolution of the densities read
\begin{equation} \begin{split}
	\frac d{dt}\braket{n_1} &= \alpha + c_0 + c_1 - (2c_0 + 2c_1 + \alpha + \gamma)\braket{n_1}\,, \\
	\frac d{dt}\braket{n_j} &= 2c_0+c_1+c_2 - (2+2c_0+2c_1+2c_2)\braket{n_j} + (2-2c_0)\braket{n_{j-1}} \ , \qquad 1 < j < L \,,\\
	\frac d{dt}\braket{n_L} &= \delta + c_0 + c_2 -(2+2c_2+\delta+\beta)\braket{n_L} + (2-2c_0)\braket{n_{L-1}}\,.
\end{split} \end{equation}

Asking for stationarity, we get
\begin{equation} \begin{split} \label{eq:dens}
	\braket{n_j} &= \frac12\Big( 1 - \Lambda \Gamma^{j-1}\Big) \ , \qquad 1 \leq j < L\,, \\
	\braket{n_L} &=  \frac12\Big( 1 - \wh\Lambda\Big) -\frac{1-c_0}{2+2c_2+\delta+\beta}\,\Lambda \Gamma^{L-2}\,.
\end{split} \end{equation}

Performing a direct diagonalisation of the Markov matrix, we checked that the above formulas correspond to the densities when $L=2,3,4$. 

\begin{figure}[htb]
\begin{center}
\includegraphics[width=\linewidth]{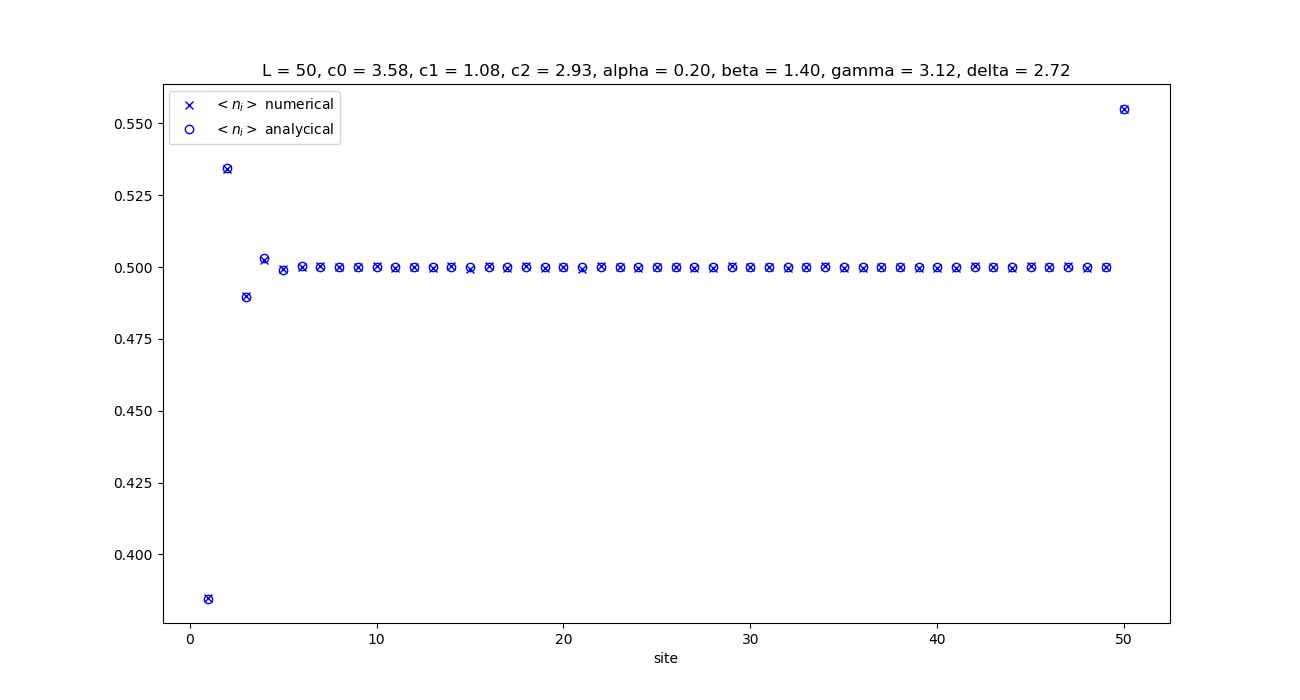}
\end{center}
\vspace{-2.4ex}
\caption{Oscillation of the densities close to the left boundary\label{fig:density_oscillo}}
\end{figure}

Remark that if one chooses the parameters $c_k$ in such a way that $\Gamma< 0$, the densities follow a damped  oscillating form. 
We illustrate it in figure \ref{fig:density_oscillo}. 

Note that, as in the SSEP case, the differential equation for the one-point correlation function decouples from the higher correlations, despite there is
evaporation and condensation. This confirms that the model with symmetric evaporation--condensation process is very similar to the SSEP.

However, although we are dealing with a SSEP-like model, which is symmetric in the bulk, the left/right symmetry is broken here by the evaporation/condensation processes. Indeed, the densities depend essentially on the left boundary, and mildly on the right one.
We checked that there is no possibility to connect the model to an ASEP model through transformations of the form \eqref{eq:conjug}.

\subsection{Currents}
From the expression of the one-point correlation function, we can deduce the currents.
There are two types of currents: the lattice current, corresponding to particles moving in the lattice, 
and the evaporation current corresponding to evaporation/condensation of particles and 'molecules' on the lattice.

\subsubsection{Lattice current and reservoir densities}
The current in the lattice reads
\begin{equation} 
\begin{split}
	J_{0,1}^\text{latt} &= \alpha \braket{1-n_1} - \gamma \braket{n_1}\,,\\	
	J_{j,j+1}^\text{latt} &= \braket{n_j(1-n_{j+1})} - \braket{(1-n_j)n_{j+1}} = \braket{n_j}-\braket{n_{j+1}}, \qquad 1 \leq j \leq L-1\,, \\
	J_{L,L+1}^\text{latt} &= \beta \braket{n_L} - \delta \braket{1-n_L}\,,
\end{split} 
\end{equation}
where  by convention $j=0$ and $j=L+1$  stand for the left and the right reservoir respectively. 
Plugging the values \eqref{eq:dens}, we get:
\begin{equation} \begin{split} \label{eq:Jlatt}
	J_{0,1}^\text{latt} & = -(c_0+c_1)\Lambda\,, \\
	J_{j,j+1}^\text{latt} &= \frac12\,\Lambda\big(\Gamma^{j}-\Gamma^{j-1}\big) 
	=-\frac{2c_0 +c_1 +c_2}{2(1+c_0+c_1+c_2)} \Lambda \Gamma^{j-1} \ , \qquad 1 \leq j \leq L-2\,, \\
	J_{L-1,L}^\text{latt} & = \frac12 \wh\Lambda -\frac{\beta+\delta+2c_0+2c_2}{2(2+2c_2+\delta+\beta)}\,\Lambda \Gamma^{L-2}\,,\\
	J_{L,L+1}^\text{latt} & =(1+c_2)\wh\Lambda- \frac{(\beta+\delta)(1-c_0)}{2+2c_2+\delta+\beta}\,\Lambda \Gamma^{L-2}\,.
\end{split} \end{equation}
Remark that generically, the currents depend only on the sum $c_1+c_2$, while it depends separately on $c_1$ and $c_2$ for the first and the two last sites.

Extending the relation $J^{latt}_{j,j+1}=\braket{n_{j}}-\braket{n_{j+1}}$ to $j=1$ and $j=L+1$ allows us to define the reservoir densities:
\begin{equation} \begin{split} 
	\braket{n_0} &= \frac 12 \Big(1-(1+2c_0+2c_1)\Lambda \Big)\,, \\
	\braket{n_{L+1}} &= \frac{\delta+(1-\beta+\delta)(1+c_2)}{2+2c_2+\delta+\beta}-\frac{(1-\beta-\delta)(1-c_0)}{2+2c_2+\delta+\beta}\,\Lambda \Gamma^{L-2}\,.
\end{split} 
\end{equation}

\subsubsection{Evaporation current}
To compute the evaporation current, one needs to consider three adjacent sites (because of the condensation of 'molecules').
For  $2\leq j\leq L-1$, it is given by 
\begin{equation} \begin{split}
	J^{ev}_{j} &= ({1+c_2-c_0})\,\Big\langle (1-n_{j-1})n_{j} - n_{j-1}(1-n_{j})\Big\rangle
+({c_2+c_0})\,\Big\langle n_{j-1}n_{j}- (1-n_{j-1})(1-n_{j})\Big\rangle \\
&\quad + ({c_1+c_0-1})\,\Big\langle n_{j}(1-n_{j+1}) - (1-n_{j})n_{j+1}\Big\rangle 
+({c_1+c_0})\,\Big\langle n_{j}n_{j+1}- (1-n_{j})(1-n_{j+1})\Big\rangle 
\\
&=  (2c_0-1)\langle n_{j-1}\rangle +2(c_2+c_1+c_0)\langle n_{j}\rangle +\langle n_{j+1}\rangle -(c_1+c_2+2c_0)\,.
\end{split} \end{equation} 

The same calculation for  the first and last sites leads to
\begin{equation} \begin{split}
	J^{ev}_{1} 
&=(2c_1+2c_0-1)\langle n_{1}\rangle+ \langle n_{2}\rangle -(c_0+c_1)\,, \\
J^{ev}_{L} 
&=(2c_0-1)\langle n_{L-1}\rangle + (2c_2+1)\langle n_{L}\rangle  -(c_2+c_0)\,.
\end{split} \end{equation}

Plugging formulas \eqref{eq:dens} in the above expressions, we get
\begin{equation} \begin{split} \label{eq:Jev}
	J_1^\text{ev} &= - \frac12 \Lambda  \left(2c_0+2c_1-1+\Gamma\right) \,,\\
	J_j^\text{ev} &= - \frac12 \Lambda  \left( 1-\Gamma \right)^2 \Gamma^{j-2} \,,\\
	J_{L-1}^\text{ev} &=- \frac12 \wh\Lambda - \left(\frac{1-c_0}{2+2c_2+\delta+\beta}+c_0+c_1+c_2\right)\Lambda \Gamma^{L-2} + \frac{1-2c_0}2 \Lambda \Gamma^{L-3} \,,\\
	J_{L}^\text{ev} &= - \frac12 (1+2c_2)\wh\Lambda+\frac{(\delta+\beta)(1-2c_0)-2(c_0+c_2)}{2(2+2c_2+\delta+\beta)}\,\Lambda \Gamma^{L-2}\,.
\end{split} \end{equation}
Using the expressions \eqref{eq:Jlatt} and \eqref{eq:Jev}, it is easy to check the current conservation on each site of the lattice:
$$
J^{latt}_{j-1,j}=J^{latt}_{j,j+1}+J^{ev}_{j} \,,\qquad 1\leq j\leq L\,.
$$

\begin{figure}[htb]
\centering
  \centering
  \includegraphics[width=\linewidth]{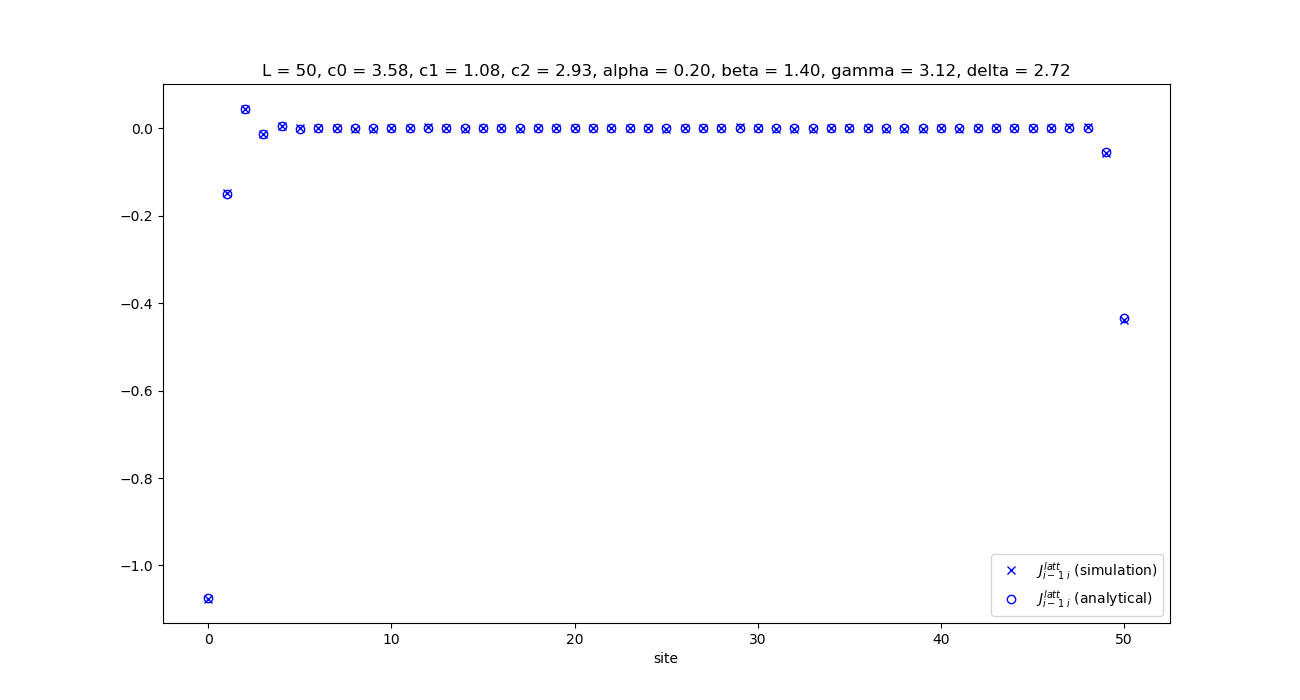}
\vspace{-2.4ex}
\caption{Lattice current \label{fig:J_latt}}

  \centering
  \includegraphics[width=\linewidth]{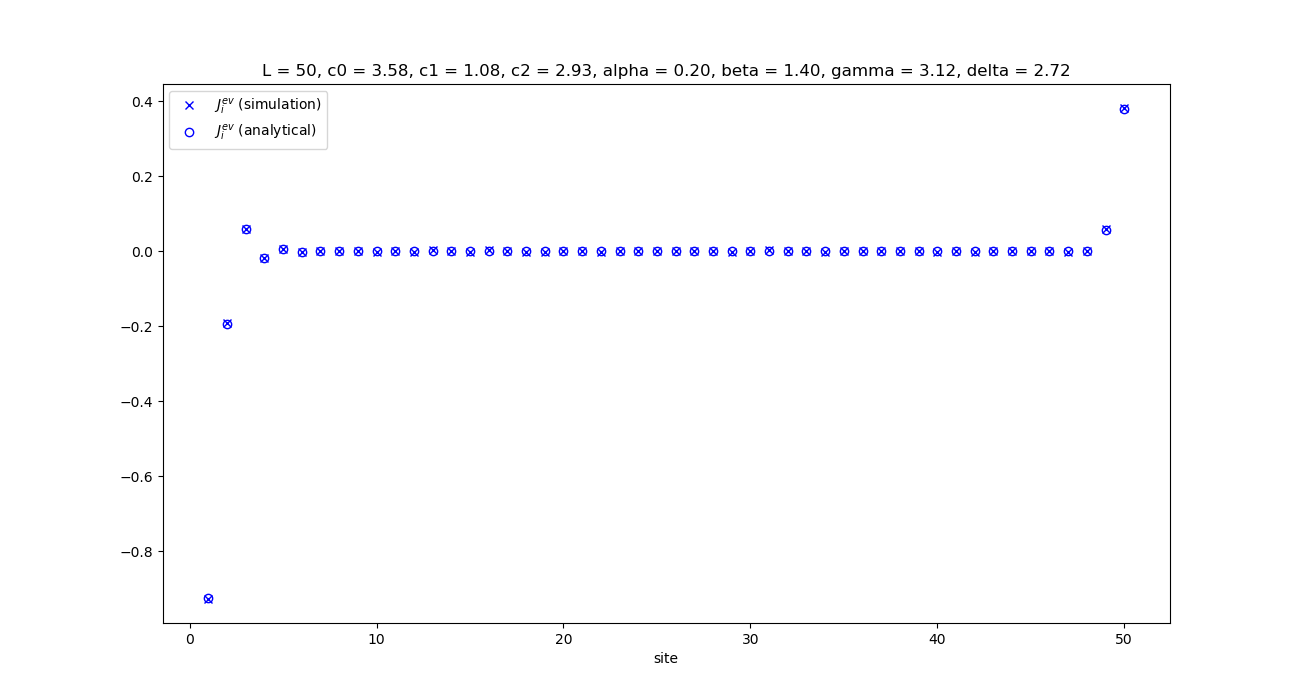}
\vspace{-2.4ex}
\caption{Evaporation current \label{fig:J_evap}}
\end{figure}

We plot below (figures \ref{fig:J_latt} and \ref{fig:J_evap}) the currents  for a system of $L=50$ sites: 
numerical results obtained from a Monte-Carlo simulation coincide with the values obtained from the analytical results.

\clearpage

\subsection{Catalysis vs fracturing processes \label{sec:fract}}
We want to compare evaporation of molecules with respect to evaporation of single particles.
Through this comparison, we can understand whether the lattice (understood as immersed in a volume, while connected to the two reservoirs) acts as a catalyst in the coagulation (resp. fracturing) reaction of particles (resp. molecules).

The current corresponding to molecule evaporation is given by
\beq
\begin{aligned}
J_{j,j+1}^{mol} &= {c_0}\,\big\langle n_{j}n_{j+1}- (1-n_{j})(1-n_{j+1})\big\rangle  =
c_0\Big(\langle n_{j}\rangle +\langle n_{j+1}\rangle -1\Big)\,,\quad 1\leq j\leq L-1.
\end{aligned}
\eeq
It leads to
\beq
\begin{aligned}
J_{j,j+1}^{mol} &= -\frac12c_0  \Lambda  (1+\Gamma)\,\Gamma^{j-1} \,,\quad 1\leq j\leq L-2
\\
J_{L-1,L}^{mol} &= -\frac12c_0 \wh\Lambda-c_0\frac{\delta+\beta+4+2c_0+2c_2}{2(2+2c_2+\delta+\beta)}\Lambda \Gamma^{L-2} 
\end{aligned}
\eeq

When $J_{j-1,j}^{mol}$ is positive, molecules evaporate from the lattice more than they condensate, whereas when $J_{j-1,j}^{mol}$ is negative, molecules mostly condensate onto the lattice. When $\Gamma < 0$, the sign of $J_{j-1,j}^{mol}$ alternates with respect to the parity of $j$.

To compare the evaporation of molecule with that of single particles, we need to consider $J^{ev}_{j-1}+J^{ev}_{j}$, 
which contains both molecules and particles evaporation (on 2 sites), and remove the part coming from molecules.
On one site, the current evaporation without molecule evaporation reads:-
\begin{equation} \begin{split}
	J_1^\text{part} &= (2c_1+c_0-1)\langle n_{1}\rangle+(1-c_0)\langle n_{2}\rangle-c_1\,, \\
	J_j^\text{part} &= (c_1+c_2)(2\braket{n_j}-1)+(1-c_0)(\braket{n_{j+1}}-\braket{n_{j-1}})\ ,\quad 2 \leq j \leq L-1\,, \\
	J_L^\text{part} &= (1-c_0+2c_2)\langle n_{L}\rangle-(1-c_0)\langle n_{L-1}\rangle-c_2 \,.
\end{split} \end{equation}

\begin{figure}[htb]
 \centering
 \includegraphics[width=\linewidth]{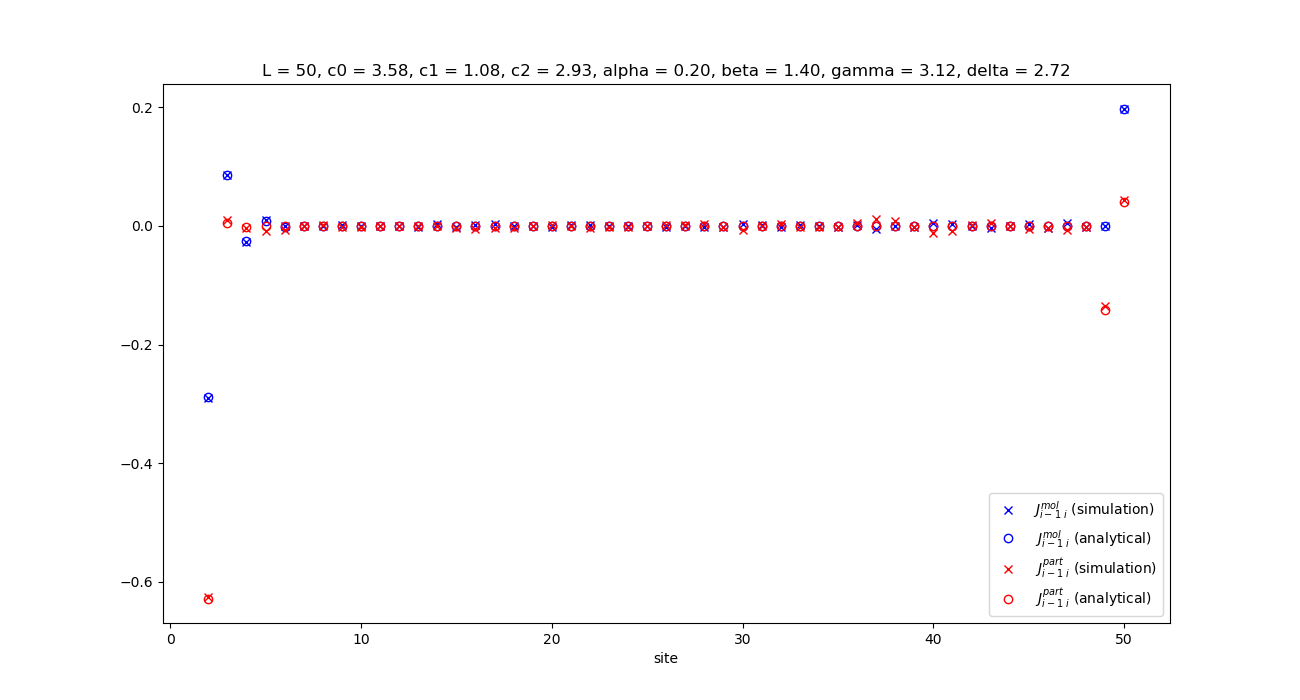}
\vspace{-2.4ex}
\caption{Evaporation currents (for molecules and particules separately) \label{fig:J_mol_part}}

\begin{center}
\includegraphics[width=\linewidth]{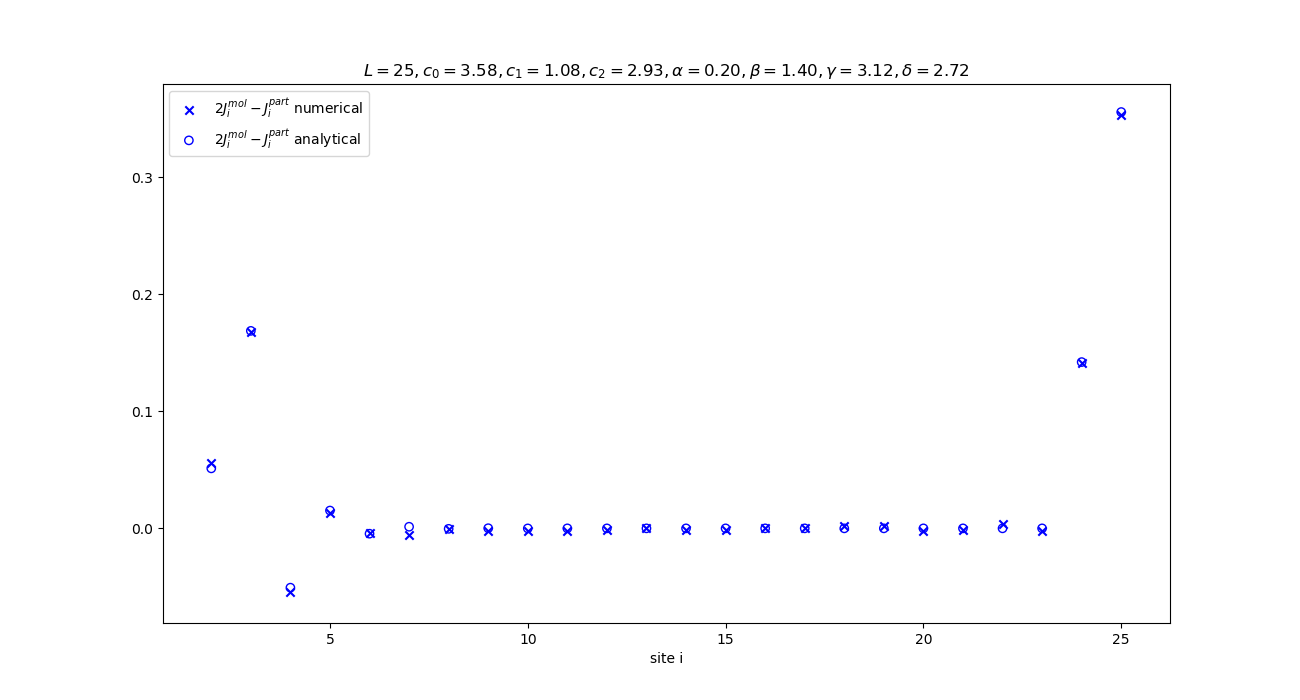}
\end{center}
\vspace{-2.4ex}
\caption{Comparison of the molecule and particle flows\label{fig:diff_mol_part}}
\end{figure}

For $2\leq j\leq L-3$, it leads to
\begin{equation} \begin{split}
	J_{j,j+1}^\text{part} &= J_{j}^\text{part} + J_{j+1}^\text{part} = 
	\Lambda \Gamma^{j-1}(1+\Gamma)\left(\frac{1-c_0}2\left(\frac 1 \Gamma - \Gamma \right) - (c_1+c_2)\right)
\end{split} \end{equation}
We show in figure \ref{fig:J_mol_part} the two parts of the evaporation current.

Remark that the ratio $\fr_{j,j+1}=J^\text{part}_{j-1,j}/J^\text{mol}_{j-1,j}$
 is constant all over the lattice except when $j \in \lbrace 1,L-2,L-1 \rbrace$:
\begin{equation} \begin{split}
	\fr_{j,j+1}&=
	\frac{(1-c_0)(\Gamma - \frac 1 \Gamma) + 2(c_1+c_2)}{c_0} =
	\frac{(c_{1}+c_{2})^2-4c_0}{c_0(1+c_0+c_{1}+c_{2})}\,,\quad 1<j<L-2\,.
\end{split} \end{equation}
It is also independent of $L$, despite the total evaporation current $J^{ev}_j$ tends to zero in the bulk when $L\to\infty$.

We define
\[
J_{tot}^{ev} = \sum_{j=1}^L J_j^{ev}\,, \qquad J_{tot}^{mol} = \sum_{j=1}^{L-1} J_{j,j+1}^{mol}\,, \qquad J_{tot}^{part} = \sum_{j=1}^L J_j^{part}\,.
\]
In the limit $L \to \infty$, the total evaporation current  simplifies, which allows us to determine the propensity of the lattice to bound particles into molecules or to split them apart.

\begin{equation} \begin{split}
	J_{\infty}^{ev}\ =\ \lim_{L\to\infty} J_{tot}^{ev} &= - (c_0+c_1)\Lambda - (1+c_2)\wh\Lambda\,, \\
	J_{\infty}^{mol}\ =\ \lim_{L\to\infty} J_{tot}^{mol} &= - \frac{c_0}2 \left(\frac{1+\Gamma}{1-\Gamma} \Lambda + \wh\Lambda \right)\,, \\
	J_{\infty}^{part}\ =\ \lim_{L\to\infty} J_{tot}^{part} &= \left( \frac{2c_0 \Gamma}{1-\Gamma} - c_1 \right) \Lambda + \left(c_0 -1 -c_2 \right) \wh\Lambda \,.
\end{split} \end{equation}

Now, the quantity 
\[ 2 J_{\infty}^{mol} - J_{\infty}^{part} = \left(c_1 - c_0 \frac{1+3\Gamma}{1-\Gamma} \right) \Lambda + (1+c_2-2c_0) \wh\Lambda\]
is positive if molecules are created by the lattice, and negative otherwise. The factor 2 before $J_{tot}^{mol}$ is necessary to compare the flow of 2-particles molecules
with the flow of single atoms. Remark that this quantity depends on the left boundary, but not on the right one.
We show in figure \ref{fig:diff_mol_part} how the local quantity $2 J^{mol}_j - J^{part}_j$ evolves with respect to the position of the site.

\subsection{Left-right asymmetry}
As already noticed, the currents and densities depends heavily  on the left boundary parameters, while right boundary has only little impact. 
This left-right asymmetry may sound surprising since the model is of SSEP-type for its diffusive aspects. 

Looking at the densities \eqref{eq:dens}, it is clear that the asymmetry is due to the presence of $\Gamma\neq0$. 
Indeed, when $c_0=1$, the symmetry between the left and right boundary is restored, see section \ref{sec:c0=1}.
Then, this asymmetry may be linked to the fact that the transposition of local Markov matrix \eqref{eq:Markovmat} is still a Markov matrix, but of a different kind.
Indeed, when $c_0=1$, the Markov matrix is self-transposed.
This suggests that the asymmetry may be induced by the evaporation/condensation processes, but a physical explanation is lacking at this point.

\clearpage

\section{$n$-point correlation functions\label{sec:correl-fct}}
The stationarity of the $n$-point correlation function leads to a recursion relation involving $\ell$-point correlation function with $\ell\leq n$ only.
In other words, we have a decoupling of the $n$-point correlation functions from the higher ones.
Thus, they can be  computed exactly. 
We illustrate it for $n=2$ in section \ref{sec:2pt-corr}, and prove the decoupling in the general case in section \ref{sec:npt-corr}. 
Let us stress that although we get an analytical expression for the correlation functions, 
they differ from the mean field: we illustrate it in section \ref{sec:num} through numerical calculations.

To simplify the presentation, we use the notations \eqref{eq:Lambda} and
\begin{equation}
\begin{aligned}
\Theta&=\frac{1-c_0}{1+c_0+2c_{1}+2c_{2}}\,,
\qquad &&\Xi = \frac{1-c_0}{1+c_0+c_{1}+c_{2}+\frac{\alpha+\gamma+\beta+\delta}2}\,,\\
\Upsilon &=\frac{1-c_0}{1+2c_0+2c_{1}+c_{2}+\frac{\alpha+\gamma}2}\,,
\qquad
&&\Delta =\frac{1-c_0}{2+c_0+c_{1}+2c_{2}+\frac{\beta+\delta}2}
\,.
\end{aligned}
\end{equation}

\subsection{Two-point correlation function\label{sec:2pt-corr}}

Considering the stationarity of the two-point correlation functions, we get the following recursion relations
for $1<i<j-1<L-1$:
\beq\label{eq:recurr_gen}
\braket{n_i n_j}  - \frac{\Gamma}{2}\Big(\braket{n_{i-1}n_j} + \braket{n_i n_{j-1}}\Big)
=\frac{1-\Gamma}4\,\Big(\braket{n_i} + \braket{n_j}\Big)\,.
\eeq
For $1<i<L-1$, we have
\bea
&&\braket{n_i n_{i+1}} - \Theta\,\braket{n_{i-1} n_{i+1}}
\ =\ \frac{(-1+c_0+c_1+c_2)\braket{n_{i+1}}+(1-c_0+c_1+c_2)\braket{n_i}+c_0 }{2+2c_0+4c_1+4c_2}\,,\qquad
 \label{eq:recurr_adj}
\\[1.ex]
&&\braket{n_1 n_{i+1}}  -  \Upsilon\,\braket{n_1n_{i}}
\ =\ 
\frac{(2c_0+c_1+c_2)\braket{n_1}+(\alpha+c_0+c_1)\braket{n_{i+1}} }{\alpha+\gamma+2+4c_0+4c_1+2c_2}\,,
\label{eq:recurr_1i}
\\[1.ex]
&&\braket{n_i n_L}  -   \Delta\,\Big(\braket{n_in_{L-1}}+\braket{n_{i-1}n_L}\Big)
\ =\  \frac{(\delta+c_0+c_2)\braket{n_i} + (2c_0+c_1+c_2) \braket{n_L}}{\beta+\delta+4+2c_0+2c_1+4c_2}\,,
\label{eq:recurr_iL}
\eea
together with the boundary cases
\beq\label{eq:n12}
\braket{n_1n_2}  \ =\  \frac{(1-c_0+c_1+c_2)\braket{n_1} + (\alpha-1+c_1)\braket{n_2}+c_0}{\alpha+\gamma+2c_0+4c_1+2c_2}\,,
\eeq
\beq
\braket{n_1 n_L}  -
\Xi\,\braket{n_1n_{L-1}}
\ =\ \frac{(\delta+c_0+c_2)\braket{n_1} +(\alpha+c_0+c_1)\braket{n_L}}{\alpha+\gamma+\beta+\delta+2+2c_0+2c_1+2c_2}\,,
\eeq
\beq\label{eq:nL1nL}
\braket{n_{L-1} n_L} - \frac{2(1-c_0)\braket{n_{L-2}n_L}}{\beta+\delta+2+2c_1+4c_2}
 \ =\  \frac{ (-1+c_0+c_1+c_2)\braket{n_L} +(\delta+1-2c_0+c_2)\braket{n_{L-1}}+c_0}{\beta+\delta+2+2c_1+4c_2}\,.
\eeq

Notice that the correlation functions $\braket{n_i n_j}$ with $j<L$ decouple from the correlation functions $\braket{n_i n_L}$, $\forall i$. 
Then, we can solve the former independently from the later.  It leads to the following expressions for the two-point correlation functions.

\begin{equation}\label{eq:solu2pt}
\begin{split}
       \braket{n_i n_j} &=
               \frac 1 4 \Bigg( 1 - (\Gamma^{i-1}+\Gamma^{j-1}) \Lambda 
               + \Gamma^{i+j-2}\Big(\frac {j-i-1} \Theta - \frac {j-i-2} \Gamma\Big) \omega \Bigg)\\
        &  + \frac14
              \sum_{k=3}^{j} \left(\frac \Gamma 2 \right)^{i+j+1-k} a_k \,\bigg\{ {i+j-k-1 \choose i-1} - {i+j-k-1 \choose j-2}\\
               & \qquad\qquad\qquad + \sum_{s=1}^{i+2-k} \left[{i+j-k-1-s \choose j-3} - {i+j-k-1-s \choose j-2} \right] \left(\frac{2 \Theta}\Gamma \right)^s \bigg\} 
               \end{split}
\end{equation}
for $1\leq i < j\leq L-1$, where
\begin{equation}
a_k = \Big(\lambda_1\,k+\lambda_0\Big)\Gamma^{k-3} -\frac{2+2c_2-\alpha-\gamma}{2(1-c_0)}(\lambda_0+\lambda_1)\Upsilon^{k-2}\,,\qquad 3 \leq k\leq L-1\,,
\end{equation}
with
\beq\begin{aligned}
\lambda_1 &=-2\,\frac{c_1+c_2}{1-c_0}\,\omega
\ ,\qquad
\lambda_0 =-2\,\frac{1+c_0}{1-c_0}\, \omega
+2\, \frac{(\alpha-\gamma)^2}{(2c_0+2c_1+\alpha+\gamma)^2}\,,
\\[1ex]
\omega &= 4 \braket{n_1n_2}-\frac{2(c_0+c_1)(1+c_0+c_1+c_2)+(\alpha+\gamma)c_0+2\alpha(c_1+c_2)+3\alpha-\gamma}{(1+c_0+c_1+c_2)(2c_0+2c_1+\alpha+\gamma)}\,.
\end{aligned}
\eeq

The expression \eqref{eq:solu2pt} is obtained as follows. The first line is a particular solution of the recurrence relations \eqref{eq:recurr_gen} and \eqref{eq:recurr_adj}, whatever the value of $\omega$. The two last lines, whatever the value of the coefficients $a_k$, solve them with vanishing r.h.s. The coefficients $a_k$ are fixed in such a way that the last relation \eqref{eq:recurr_1i}   is also obeyed and $\omega$ is given by the initial value \eqref{eq:n12}. 
This allows to get all two-point functions $\braket{n_in_j}$ with $1\leq i<j<L$.

The remaining two-point functions are more involved. We get
\begin{equation}
\begin{aligned}
\braket{n_1 n_L} &= \frac{\alpha+c_0+c_1}{(2c_0+2c_1+\alpha+\gamma)(2+2c_2+\beta+\delta)}\Big(
1+c_2+\delta-(1-c_0)\,\Lambda\, \Gamma^{L-2}\Big) \\[1ex]
&\quad- \frac18(\lambda_0+\lambda_1)\,\Gamma\,\Xi\,\Upsilon^{L-3}\,,
\end{aligned}
\end{equation}
and for $2 \leq i \leq L-2$,
\begin{equation}
\begin{aligned}
\braket{n_i n_L} &= \Delta^{i-1}\braket{n_1 n_L} +\sum_{k=1}^{i-1} \Delta^{i-k}\braket{n_{k+1} n_{L-1}}
-\frac14 \frac{\delta+c_0+c_2}{1-c_0}\Lambda\, \frac{\Delta\Gamma^i-\Gamma\Delta^{i}}{\Gamma-\Delta}\\
&\quad+\frac{\Delta-\Delta^{i}}{\Delta-1}\left( \frac{\delta+c_0+c_2}{4(1-c_0)}+\frac{2c_0+c_1+c_2}{2(1-c_0)}(1-\wh{\Lambda})
-\frac{1-c_0}{\delta+\beta+2+2c_2}\Lambda\Gamma^{L-2}\right)\,.
\end{aligned}
\end{equation}
Finally $\braket{n_{L-1} n_L}$ is given by formula \eqref{eq:nL1nL}.

\subsection{Hierarchical relations for $n$-point correlation function\label{sec:npt-corr}}
As for the two-point correlation functions, the $n$-point correlation functions decouple from the higher ones, and thus can be exactly computed recursively.
To show it, we introduce the following row vectors
\beq
\texttt u=(1,1)\,,\quad \texttt f=(0,1), \quad
\texttt v_n(i_1,\dots,i_n) = \texttt u\otimes \dots \otimes \texttt u\otimes\underbrace{\texttt f}_{i_1}\otimes\,
 \texttt u \dots \texttt u\otimes\underbrace{\texttt f}_{i_n}\otimes\, \texttt u \dots \texttt u
\eeq
so that the $n$-point correlation functions read $\braket{n_{i_1}\dots n_{i_n}}=\texttt v_n(i_1,\dots,i_n)|\cS\rangle$.

The time evolution of this correlation function is ruled by the action of the Markov matrix $\MM$ on 
$\texttt v_n(i_1,\dots,i_n)$. Thus, the recursion relations for the correlation functions are obtained from the identities
\beq
\begin{aligned}
\texttt u\otimes \texttt u\, M &= 0 \mb{;} \texttt u\, B = 0\mb{;} \texttt u\, \bar B = 0\\
\texttt u\otimes \texttt f M &= (c_0+c_2) \texttt u \otimes \texttt u\ - 2(1+c_2) \texttt u \otimes \texttt f\ + 2(1-c_0) \texttt f \otimes \texttt u\\
\texttt f\otimes \texttt u\, M &= (c_0+c_1) \left( \texttt u \otimes \texttt u\ - 2\ \texttt f \otimes \texttt u \right) \\
\texttt f \otimes \texttt f M &= c_0\ \texttt u \otimes \texttt u\ + (c_1-1) \texttt u \otimes \texttt f + (1-2c_0+c_2) \texttt f \otimes \texttt u -2(c_1+c_2) \texttt f \otimes \texttt f \\
\texttt f B &= \alpha\ \texttt u -(\alpha+\gamma)\,\texttt f\\
\texttt f \bar B &= \delta\ \texttt u -(\delta+\beta)\,\texttt f
\end{aligned}
\eeq
Since $\texttt f\otimes \texttt f$ occurs only in the action of $M$ on $\texttt f\otimes \texttt f$ (and because 
$\texttt u\otimes \texttt u M = 0$) this proves that the recursion relation for the $n$-point functions involves only 
correlations of $n$-point at most.

\subsection{$n$-point correlations recursion relations}
The aim of this section is to compute explicitly the recursion relations obeyed by the $n$-point correlation fonctions.
Let $T = \lbrace i_1 \cdots i_n \rbrace$ the set of sites where we want to calculate the correlation, i.e.
\beq
\braket{n_T} = \braket{n_{i_1} \cdots n_{i_n}}. 
\eeq
We decompose $T$ as a union of disjoint segments:
\beq
T=\cup_{a=1}^N\, [i_a\,,\,j_a] \mb{with} j_{a-1}+1< i_{a}\leq j_{a}<i_{a+1}-1 \,,\ a=1...,N\,,
\eeq
where by convention $j_{0}=-1$ and $i_{N+1}=L+2$.

The recursion relations are obtained through the action of the Markov matrix
\beq
0=\MM\,\braket{n_T}=\sum_{a=1}^{N} M_{[i_a\,,\,j_a]}\, \braket{n_T} +\delta_{i_1,1}\,B_1\, \braket{n_T}+ \delta_{j_N,L}\,\bar B_L\, \braket{n_T}\,,
\eeq
where $M_{[i_a\,,\,j_a]}=\oplus_{k=i_a}^{j_a-1} M_{k,k+1}$. It is clear that there is a decoupling of the different segments $[i_a\,,\,j_a]$ and that the contributions of each segment will add up with no interference. Then, it is sufficient to consider the sets $T=[i,j]$ which appear as building blocks for the calculation of the recursion relations.

We obtain, for $T=[i,j]$ with $1\leq i\leq j\leq L$:
\beq\label{eq:npt}
\begin{array}{c}
\Big(2(j+1-i)(c_1+c_2)+2(1+c_0) +\delta_{i,1}\,\big(\alpha+\gamma-2-2c_2\big)+\delta_{j,L}\,\big(\beta+\delta-2c_0-2c_1\big)\Big)\, \braket{n_{[i,j]}} \\
-2(1-c_0)(1-\delta_{i,1})\,\braket{n_{i-1}n_{[i+1,j]}} \\
\qquad=\\
 \Big(c_0+c_2+(1-\delta_{i,j})\, (c_1-1)+\delta_{i,1}\,\big(\alpha-c_0-c_2\big)\Big)\,\braket{n_{[i+1,j]} } \\
+\Big(c_0+c_1+(1-\delta_{i,j})\, (1-2c_0+c_2)+\delta_{j,L}\,\big(\delta-c_1-c_0\big)\Big)\,\braket{n_{[i,j-1]}} \\
\displaystyle+ c_0\sum_{k=i}^{j-1} \braket{n_{[i,k-1]}n_{[k+2,j]}} +(c_1+c_2-2c_0)\,\sum_{k=i+1}^{j-1} \braket{n_{[i,k-1]}n_{[k+1,j]}} \,,
\end{array}
\eeq
where we have used the conventions that $n_{[i\,,\,i]}=n_{i}$, $n_{[k\,,\,\ell]}=1$ when $k>\ell$, and $\sum_{k=a}^b(\cdots)=0$ if $a>b$.

In \eqref{eq:npt}, the left-hand-side corresponds to $(j+1-i)$-point correlation functions, while the right-hand-side involves only $m$-point correlation functions with $m\leq j-i$.
Since for $i=1$, the l.h.s. contains only the correlation $ \braket{n_{[1,j]}}$, it can be solved by recursion hypothesis. Then, the correlation $ \braket{n_{[2,j+1]}}$ can be computed once one knows $ \braket{n_{[1,j]}}$ and the correlations with a lower number of points, and so on.
It is the left-right asymmetry of the model that allows the calculation through this (double) recursion.

When $T=\cup_{a=1}^N\, [i_a\,,\,j_a]$, the calculation is more involved, but follows also a recursion: one first needs to compute all recursions $ \braket{n_{[1,k]}}$ with 
$k\leq |T|$ following the method given above. Then, similarly, one can compute $ \braket{n_{[1,k]}n_\ell}$ starting from $\ell=k+2$, and so on.

\section{The special case $c_0=1$ \label{sec:c0=1}}
As already mentioned, the comparison with the DISSEP model is obtained for $c_0=1$ and $c_1=c_2=0$ and corresponds to the free fermions point. 
Here, we consider a more general case by setting $c_0=1$ but keeping $c_1$ and $c_2$ free.
Then we have $\Gamma=0$ and expressions simplify. The bulk Markov matrix reads
\beq
M\big|_{c_0=1}=\left(\begin{array}{cccc}
-1- c_1-c_1 & c_2 & c_1 & 1 \\ 
c_2 & -1- c_1-c_2 & 1 & c_1\\ 
c_1 & 1 & -1- c_1-c_2 & c_2 \\ 
1 & c_1 &  c_2 &-1- c_1-c_2
\end{array}\right)\,.
\eeq

The densities take the form
\beq\begin{array}{c}
\displaystyle\langle n_{0}\rangle = \frac{\alpha+(1+\alpha-\gamma)(1+c_1)}{2+2c_1+\alpha+\gamma}\quad\mb{;}\quad
\displaystyle\langle n_{1}\rangle=\frac{\alpha+1+c_1}{2+2c_1+\alpha+\gamma}\,; \\[2.4ex]
\displaystyle\langle n_{j}\rangle=\frac12\,,\quad 2\leq j\leq L-1\,; \\
\displaystyle \langle n_{L}\rangle=\frac{\delta+1+c_2}{2+2c_2+\beta+\delta}\quad\mb{;}\quad
\displaystyle\langle n_{L+1}\rangle = \frac{\delta+(1+\delta-\beta)(1+c_2)}{2+2c_2+\delta+\beta}\,.
\end{array}
\eeq
For the current in the lattice, we have
\beq
\begin{array}{c}
\displaystyle J^{latt}_{01}=\frac{(\alpha-\gamma)(1+c_1)}{2+2c_1+\alpha+\gamma} 
\qquad\qquad;\quad\qquad
\displaystyle J^{latt}_{12}=\frac{\alpha-\gamma}{2(2+2c_1+\alpha+\gamma)} \ ;
\\[4.ex]
\displaystyle J^{latt}_{j,j+1}=0\,,\quad 2\leq j\leq L-2\ ;
\\[2.2ex]
\displaystyle J^{latt}_{L-1,L}=\frac{\beta-\delta}{2(2+2c_2+\beta+\delta)} 
\qquad;\qquad
\displaystyle J^{latt}_{L,L+1}=\frac{(\beta-\delta)(1+c_2)}{2+2c_2+\beta+\delta} \,.
\end{array}
\eeq
Finally, the evaporation current reads
\beq\begin{array}{c}
\displaystyle J^{ev}_{1}=\frac{(\alpha-\gamma)(1+2c_1)}{2(2+2c_1+\alpha+\gamma)} 
\qquad;\qquad
\displaystyle J^{ev}_{2}=\frac{\alpha-\gamma}{2(2+2c_1+\alpha+\gamma)}\ ;
\\[4ex]
 \displaystyle J^{ev}_{j}=0\,,\quad 3\leq j\leq L-2\ ;
\\[2.1ex]
\displaystyle J^{ev}_{L-1}=\frac{\delta-\beta}{2(2+2c_2+\beta+\delta)}
\qquad;\qquad
\displaystyle J^{ev}_{L}=\frac{(\delta-\beta)(1+2c_2)}{2(2+2c_2+\beta+\delta)}\,.
\end{array}
\eeq
When $c_1=c_2=0$, these values agree with the ones computed in \cite{DiSSEP}.

Remark that we also have
\beq
J^{part}_{j}= -2(c_1+c_2)(\braket{n_{j}}+\braket{n_{j+1}}-1)=0 \mb{and}{J^{mol}_{j,j+1}} = \braket{n_{j}}+\braket{n_{j+1}}-1=0\,,
\eeq
which shows that in the limit $c_0\to1$, there is an opposite flux of condensation/evaporation for particules and for molecules, 
the ratio being constant and equal to $-2(c_1+c_2)$. 
The competition between the two processes is parametrized by $c_1+c_2$, which can be adjusted to favor one or the other.

In fact, when $c_0=1$, the bulk Markov matrix is diagonalizable:
\beq\label{eq:Mdiago}
(U\otimes U) \, M\big|_{c_0=1}\, (U^{-1}\otimes U^{-1}) 
=\left(\begin{array}{cccc}
0 & 0 & 0 & 0 \\ 
0 & -2(1+c_2) & 0 & 0\\ 
0 & 0 & -2(1+c_1) & 0 \\ 
0 & 0 &  0 &-2(c_1+c_2)
\end{array}\right),
\eeq
where 
$$U=\begin{pmatrix} 1 & 1 \\ 1 & -1\end{pmatrix}.$$
Performing this transformation for the full Markov matrix \eqref{eq:marko-full}
$$
\MM\big|_{c_0=1}\,\to\, U\otimes U\otimes \cdots \otimes U\,\MM\big|_{c_0=1}\,U^{-1}\otimes U^{-1}\otimes\cdots \otimes U^{-1}
\,,
$$
it is easy to see that for the transformed Markov matrix, the  eigenstate with zero eigenvalue reads 
\beq
|0\rangle = \begin{pmatrix} 1\\ \Lambda\end{pmatrix}\otimes  \begin{pmatrix} 1\\ 0\end{pmatrix}\otimes 
\cdots\otimes  \begin{pmatrix} 1\\ 0\end{pmatrix}\otimes  \begin{pmatrix} 1\\  \wh\Lambda\end{pmatrix}\,,
\eeq
where $\Lambda$ and $\wh\Lambda$ are given in \eqref{eq:Lambda} (with $c_0=1$).
This allows us to compute the stationary state of the Markov matrix $\MM$:
\beq\label{eq:stat-state-c0}
|{\cS}\rangle = U^{-1}\otimes U^{-1}\otimes\cdots \otimes U^{-1}\,|0\rangle=
\begin{pmatrix} 1+ \Lambda\\ 1- \Lambda\end{pmatrix}\otimes  \begin{pmatrix} 1\\ 1\end{pmatrix}\otimes 
\cdots\otimes  \begin{pmatrix} 1\\ 1\end{pmatrix}\otimes  \begin{pmatrix} 1+\wh\Lambda\\  1-\wh\Lambda\end{pmatrix}.
\eeq
Remark that the symmetry between the left and the right boundaries has been re-established.

Then, the calculation of the correlation functions becomes easy, and one can show that we get the results given above. 
For the two-point functions, one can check directly that the 
mean field approximation becomes exact at $c_0=1$:
$$
\langle n_i\,n_j\rangle = \langle n_i\rangle\,\langle n_j\rangle\,,\quad 1\leq i\neq j\leq L.
$$
As we show in section \ref{sec:num} below,  it is not true for generic values of $c_0$.

\section{Numerical calculations\label{sec:num}}

To confirm our analysis, we have performed numerical calculations from two different methods. The first method consists in solving numerically the linear system which must satisfy the stationary state, \textit{i.e.} finding the normalized vector spanning the kernel of the Markov matrix. We used this method to get really accurate results up to $L=14$ sites. For larger values of $L$, we used a Monte-Carlo approach.

\begin{figure}[htb]
  \centering
\vspace{-1.7ex}
  \includegraphics[width=\linewidth]{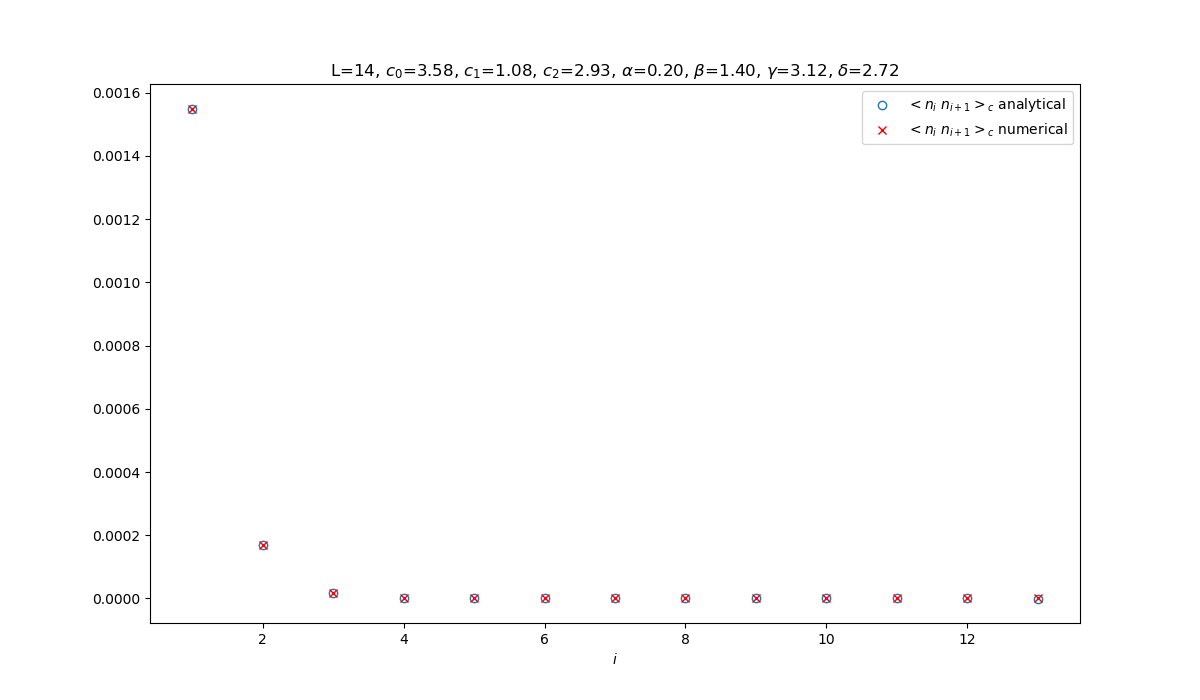}
\caption{Connected 2-point correlation $\braket{n_i n_{i+1}}_c$ \label{fig:n2c_neighbour}}

  \centering
\vspace{-0.4ex}
  \includegraphics[width=\linewidth]{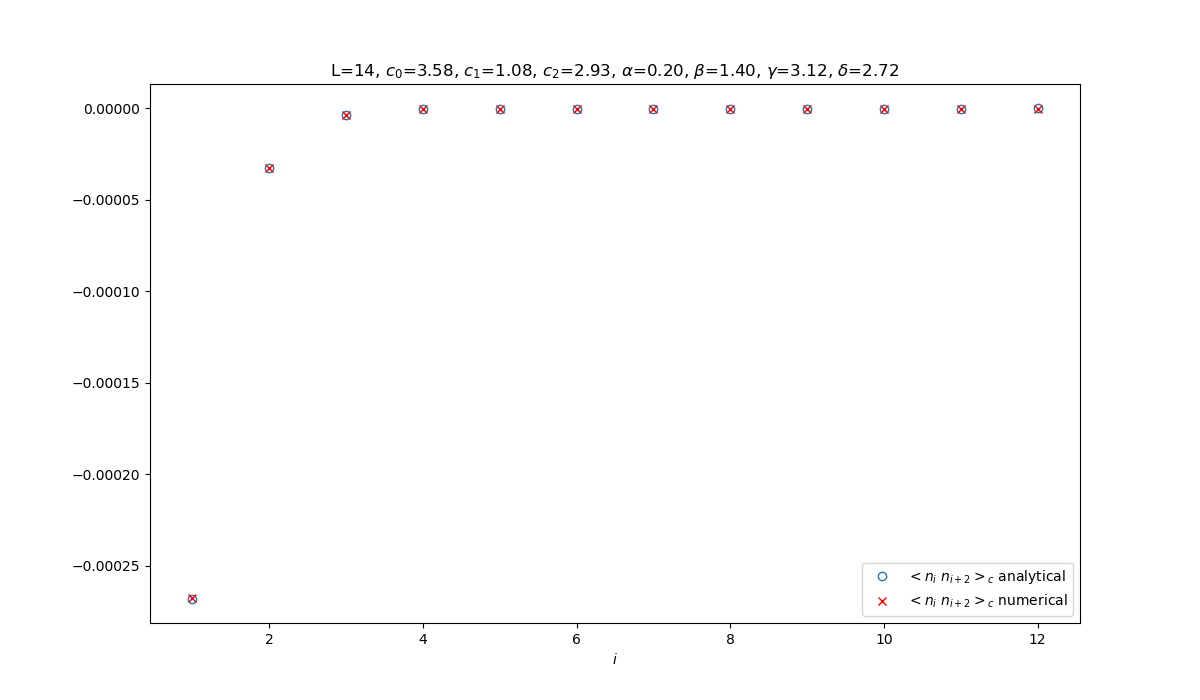}
\caption{Connected 2-point correlation $\braket{n_i n_{i+2}}_c$ \label{fig:n2c_neighbour2}}
\end{figure}

\begin{figure}[htb]
  \centering
\vspace{-1.7ex}
  \includegraphics[width=\linewidth]{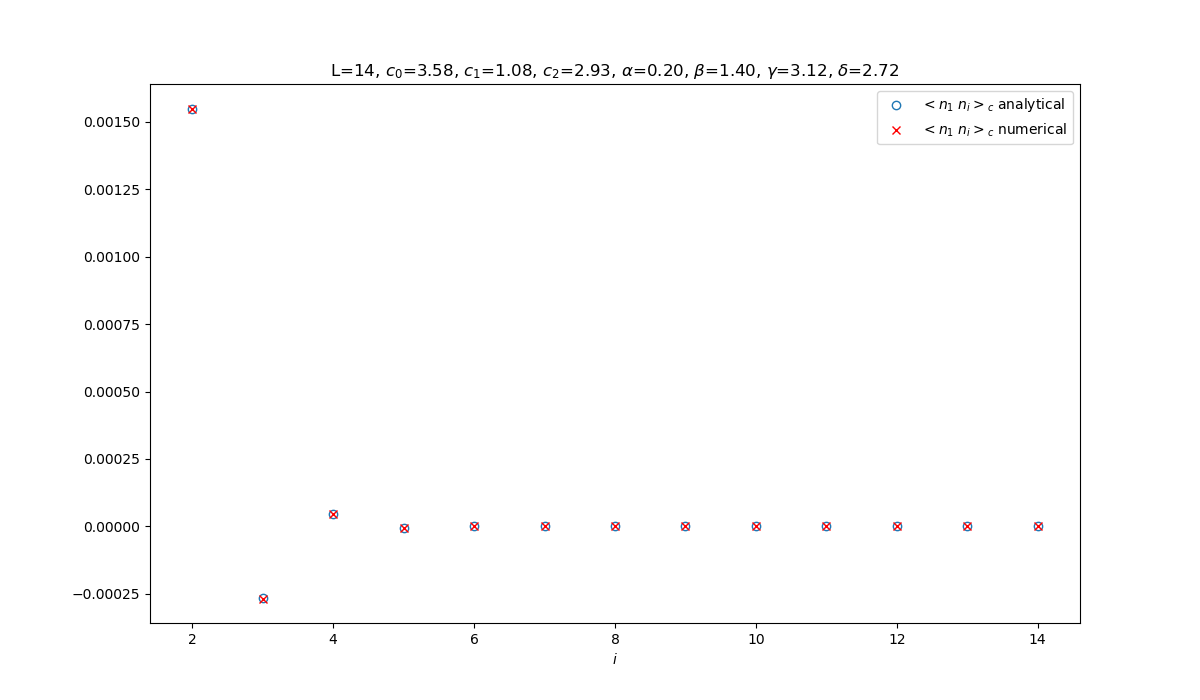}
\caption{Connected 2-point correlation $\braket{n_1 n_i}_c$ \label{fig:n2c_first}}

\centering
\vspace{-0.4ex}
  \includegraphics[width=\linewidth]{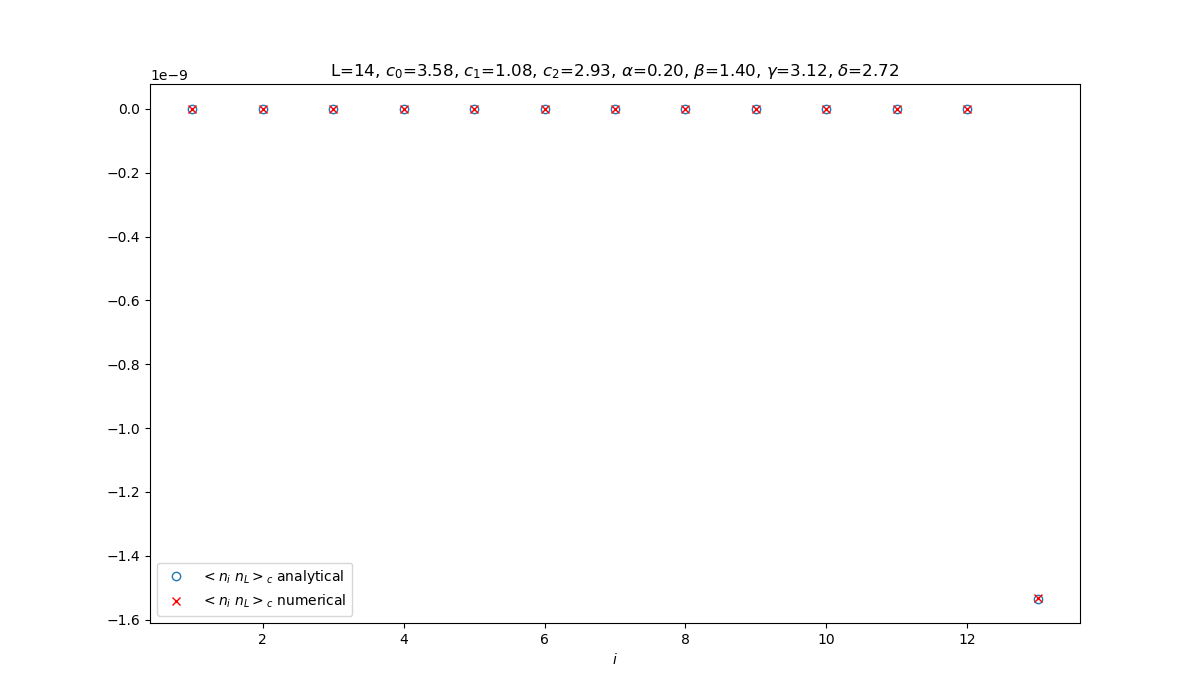}
\caption{Connected 2-points correlation $\braket{n_i n_L}_c$ \label{fig:n2c_last}}
\end{figure}

We have  computed numerically (figures \ref{fig:n2c_neighbour},  \ref{fig:n2c_neighbour2},  \ref{fig:n2c_first}, and \ref{fig:n2c_last}) the connected two-point correlation $\langle n_jn_k\rangle_c=\langle n_jn_k\rangle-\langle n_j\rangle\langle n_k\rangle$ up to $L=14$ sites by solving numerically the linear system (first method), and compared it against our analytical results.


\clearpage

\section{Conclusion}
The model we have studied possesses several features that are worth mentioning. The left/right asymmetry, in particular, is a striking point. 
Although it is clear that this asymmetry roots in the evaporation/condensation part of the model, a deeper understanding of this property needs to be developed.
Although the thermodynamical limit $L\to\infty$ seems to be trivial (most densities reach $\frac12$ and most of the currents are zero), 
the model being symmetric in its diffusive part, it is tempting to look for a WASEP-like model in the thermodynamical limit. 
We hope to come back on this point in the next future. A generalization of the model to the case of asymmetric diffusion rates is also a line that we wish to develop.

We have already mentioned that such hierarchical relations were studied also in the literature, see e.g. \cite{Klich, Eisler, GKR}, but the models where based on fermions, and/or dealt with discrete time or periodic cases. The present model does not fall in these classes.

As already mentioned, the model can be solved at the level of correlation functions, but there is no known integrable structure. 
Nevertheless, since all correlation functions can be computed analytically, a more detailed search of its possible integrability is worth doing. We have checked that  for generic values of the parameters, the matrix $M$ does not obey a braid relation.
This prevents it from being part of a Hecke, a Temperley-Lieb or a BMW algebra that would lead to a baxterisation procedure 
\cite{Jones,Jim,BMW}. However, there are several types of new algebras that can be used for a baxterisation, see
for instance the new algebras introduced in \cite{newB,back}. It would be interesting to look for a four dimensional representation of them that could lead to the Markov matrix of the present paper.

The search of a matrix ansatz is also a direction that one could look at. It seems however that the associated algebra, if it exists, is not a 'diffusion algebra', as they were 
 studied in \cite{IPR,EssRitt}. Obviously, if the model appears to be integrable, a search for a matrix ansatz would be greatly simplified, following the lines given in \cite{CRV}.

Finally, the fracturing / catalysis approach is also a point that needs to be explored. A qualitative comparison   with a true catalysis reaction in one dimension would be a first step to see whether the model can be used as a toy model for such processes.

\subsection*{Acknowledgements}
We are grateful to E. Bertin, N. Cramp\'e and V. Lecomte for fruitful discussions and comments during the course of this work.

\end{document}